\documentclass[final,3p,times,12pt]{elsarticle}

\usepackage{amsmath,amssymb}
\usepackage{graphicx}
\usepackage{rotating}
\usepackage{color}
\usepackage{xcolor}
\usepackage{fontenc}
\usepackage{slashed}
\usepackage{longtable}
\usepackage{dcolumn}
\usepackage{bm}
\usepackage{natbib}
\usepackage{lineno,hyperref}
\hypersetup{colorlinks,bookmarksopen,bookmarksnumbered,linkcolor=blue,pdfstartview=FitH,urlcolor=blue,citecolor=blue}

\journal{arXiv}

\begin{document}

\begin{frontmatter}
\title{Deviations to $\mu-\tau$ symmetry in a cobimaximal scenario}

\author[1]{Diana C. Rivera-Agudelo} 
\ead{diana.rivera11@usc.edu.co}
\address[1]{Universidad Santiago de Cali, Facultad de Ciencias B\'asicas, Campus Pampalinda, Calle 5 No. 62-00, C\'odigo Postal 76001, Santiago de Cali, Colombia}

\author[1]{S. L. Tostado}
\ead{sergio.tostado00@usc.edu.co}

\begin{abstract}
Within the different patterns of the neutrino mixing matrix, the cobimaximal
mixing remains a plausible possibility for understanding the flavor structure
of neutrinos as it is consistent with current experimental data. Such a
pattern is related to a concrete form of the mass matrix, displaying a
$\mu-\tau$ reflection symmetry, which has motivated many theoretical
investigations in recent years. In this paper, we discuss the effects of
the Majorana phases over a $\mu-\tau$ symmetric mass matrix obtained
from a cobimaximal mixing matrix. We investigate how these phases could
be restricted through deviations from the symmetric scenario and the most
precise determinations of the mixing parameters. Some of our relations
could be tested with future results.
\end{abstract}
\end{frontmatter}
%

\section{Introduction}

\label{sec:intro}

Neutrino mixing phenomena are resumed in the Pontecorvo-Maki-Nakagawa-Sakata
(PMNS) matrix \cite{Pontecorvo:1957cp,Maki:1962mu}. Global fits have reached
the percent level in determining the mixing angles, confirming the massive
nature of neutrinos. Also, there seem to be some hints leading to a non-zero
$CP$ phase
\cite
{ParticleDataGroup:2022pth,Capozzi:2018ubv,deSalas:2020pgw,Esteban:2020cvm,Gonzalez-Garcia:2021dve,Abi:2018dnh},
but no major information can be obtained for the Majorana $CP$ phases in
oscillation experiments. Nevertheless, neutrinoless double beta decay
experiments
could be sensitive to the Majorana phases \cite{Dolinski:2019nrj}, which,
in addition, could help elucidating the neutrino mass spectrum.

In the standard form, the PMNS matrix can be written in the following way
%
\begin{eqnarray}
\label{PMNS}
U_{\mathrm{PMNS}} &=& \left(
\begin{array}{c@{\quad}c@{\quad}c}
c_{12} c_{13} & s_{12}c_{13} & s_{13} e^{-i\delta_{CP}}
\\
- s_{12} c_{23} + c_{12}s_{23}s_{13}e^{i\delta_{CP}} & c_{12}c_{23} +
s_{12}s_{23}s_{13}e^{i\delta_{CP}} & -s_{23}c_{13}
\\
-s_{12}s_{23} - c_{12}c_{23}s_{13}e^{i\delta_{CP}} & c_{12}s_{23}
-c_{23}s_{12}s_{13}e^{i
\delta_{CP}} & c_{23}c_{13}
\end{array}
\right)
\nonumber
\\
&&\times{\mathrm{diag}} \left[ 1, e^{-i\frac{\beta_{1}}{2}}, e^{-i
\frac{\beta_{2}}{2}}\right]~,
\end{eqnarray}
in the case of Majorana neutrinos. Here, $s_{ij}$ and $c_{ij}$ stand for
$\sin\theta_{ij}$ and $\cos\theta_{ij}$, respectively, with
$\theta_{ij}$ denoting the mixing angles $\theta_{12}$,
$\theta_{13}$, and $\theta_{23}$. The $CP$ phases are then represented
by $\delta_{CP}$, $\beta_{1}$ and $\beta_{2}$, for Dirac and Majorana
phases, respectively.

Within the theoretical efforts, a co-Bimaximal ($CBM$) mixing matrix has
been motivated in the search of a symmetric structure in the masses and
mixings of neutrinos
\cite{Fukuura:1999ze,Miura:2000sx,Harrison:2002et,Ma:2015fpa}. This form
of the mixing matrix is obtained by considering
$\theta_{23} = \pi/4$ and $\delta_{CP} = -\pi/2$ in Eq.~(\ref{PMNS}),
in addition with $\beta_{1,2} = 0$, \textit{i.e.}
%
\begin{equation}
\label{CBM}
U_{CBM} = \left(
\begin{array}{c@{\quad}c@{\quad}c}
c_{12}~c_{13} & s_{12}~c_{13} & -i~s_{13}
\\
\frac{-1}{\sqrt{2}}(s_{12} - i~c_{12} ~s_{13}) & \frac{1}{\sqrt{2}}(c_{12}
+ i~s_{12}~s_{13}) & \frac{- c_{13}}{\sqrt{2}}
\\
-\frac{1}{\sqrt{2}}(s_{12} + i~c_{12}~s_{13}) & \frac{1}{\sqrt{2}}(c_{12}
- i~s_{12}~s_{13}) & \frac{c_{13}}{\sqrt{2}}
\end{array}
\right).
\end{equation}
Such a simple form is consistent with current determinations of
$\theta_{23}$ and $\delta_{CP}$ within $3\sigma$ error intervals, leaving
$\theta_{13}$ and $\theta_{12}$ free of taking the experimental values.
We can observe that the second and third rows in Eq.~(\ref{CBM}) display
the following relation $|U_{\mu i}| = |U_{\tau i}|$, also known as a
$\mu-\tau$ symmetry. The neutrino mass matrix can be obtained from
$M_{\nu}^{0}=U_{CBM}{\mathrm{diag}}(m_{1},m_{2},m_{3}) U_{CBM}^{
\mathrm{T}}$, and can be resumed as \cite{Ma:2017moj}
%
\begin{equation}
\label{eq:massmatrix}
M_{\nu}^{0} = \left(
\begin{array}{c@{\quad}c@{\quad}c}
m_{ee} & m_{e\mu} & m_{e\mu}^{*}
\\
m_{e\mu} & m_{\mu\mu} & m_{\mu\tau}
\\
m_{e\mu}^{*}& m_{\mu\tau} & m_{\mu\mu}^{*}
\end{array}
\right),
\end{equation}
which presents a reflection symmetry between the $\mu$ and $\tau$ labels.
Such a symmetric form is then an imprint of considering a $CBM$ mixing
matrix in computing $M_{\nu}^{0}$, and could be considered as a starting
point of any improved model of neutrinos masses and mixings, where the
effective Majorana mass term remains invariant under the partially flavor
changing $CP$ transformations
%
\begin{equation}
\nu_{eL} \rightarrow(\nu_{eL})^{c} ,~~~~~ \nu_{\mu L}
\rightarrow(\nu_{\tau L})^{c} ,~~~~~ \nu_{\tau L} \rightarrow(
\nu_{\mu L})^{c}~,
\end{equation}
where $(\nu_{\alpha L})^{c}$ stands for the charge conjugate of
$\nu_{\alpha L}$. Theoretically, the origin of this pattern may be diverse.
It could be related to a non-abelian discrete symmetry, type $A_{4}$, present
at high energies, which should be broken at small energies to match as
possible the experimental data (see for instance \cite{Ma:2015fpa}, and
references therein). Furthermore, some corrections to this pattern have
also been considered previously\footnote{Such corrections may arise, for
example, from the renormalization group equations and/or from the charged
lepton sector.} to generate $\delta_{CP}$ and $\theta_{23}$ out of the
maximal values \cite{Ma:2017moj,Zhao:2017yvw}. In most cases, any correction
imposed to $U_{CMB}$ would have direct consequences on the symmetric structure
of $M_{\nu}^{0}$.

In this work, we follow a different approach by considering a $CBM$ mixing
matrix with Majorana $CP$ phases included in the standard form. We should
expect that such phases may modify, or break, the symmetry of the mass
matrix without modifying the mixing angles and the Dirac phase. Hence,
in Sec.~\ref{sec:deviations}, we investigate the effects of the Majorana
phases on the entries of the neutrino mass matrix that break the underlying
symmetry. We modulate the deviations from the symmetric structure with
two perturbation parameters which could impose some restrictions on the
Majorana $CP$ phases. In Sec.~\ref{sec:results}, we follow a phenomenological
approach to restrict the Majorana phases and study their effects in some
physical observables. Finally, in Sec.~\ref{sec:conclusions} we present
our main conclusions.

\section{Cobimaximal mixing and CP phases}
\label{sec:deviations}

Motivated by the experimental values of the mixing angles, let us assume
in the following a $PMNS$ mixing matrix of the form
%
\begin{equation}
\label{eq:newpmns}
U_{PMNS} = U_{CBM} \times{\mathrm{diag}} \left[ 1, e^{-i
\frac{\beta_{1}}{2}}, e^{-i\frac{\beta_{2}}{2}}\right]~,
\end{equation}
where $U_{CBM}$ is of the form in Eq.~(\ref{CBM}), and $\beta_{1}$ and
$\beta_{2}$ the Majorana phases. The advantage of taking this parametrization
is that it fulfills the current values of $\theta_{23}$ and
$\delta_{CP}$ within $3\sigma$ or less, while $\theta_{12}$ and
$\theta_{13}$ are free parameters, such that they can be easily accommodated
within the $1\sigma$ experimental interval.

As can be expected, this form of the $PMNS$ matrix allows the mass matrix
to deviate from the symmetric pattern observed in Eq.~(\ref{eq:massmatrix}),
which is directly induced by the Majorana phases. It should be clear that
in the limit $\beta_{1,2}=0$, the symmetric pattern is restored.

Previous analysis \cite{Gupta:2013it,Rivera1} has shown, in the case of
a $\mu-\tau$ permutation symmetry, that departures from a symmetric mass
matrix can be resumed by two parameters that measure the breaking of the
equalities $m_{e\tau}= m_{e \mu}$ and $m_{\tau\tau}= m_{\mu\mu}$. In
particular, some corrections to the Tri-Bi-Maximal\footnote{The corrections
to the Tri-Bi-Maximal mixing usually include additional mixing angles and/or
$CP$ phases.} mixing matrix have been adopted
\cite{Rivera2,Rivera-Agudelo:2019seg,Rivera-Agudelo:2020orh}, where the
two breaking parameters of the $\mu-\tau$ permutation symmetry are related
to correction angles. Then, we should anticipate that in the $CBM$ case,
we could follow a similar approach since the inclusion of the Majorana
phases would contribute to spoiling the reflection symmetry of the mass
matrix, which, to the best of our knowledge, has not been considered previously.
However, deepen on the origin of these phases is out of the scope of the
present work.

Let us then modulate the deviations from a symmetric mass matrix in the
form
%
\begin{equation}
\label{eq:corrected}
M_{\nu}= M_{\nu}^{0} + \delta M \left(\delta,\epsilon\right)~,
\end{equation}
where $M_{\nu}^{0}$ follows the symmetric form in Eq.~(\ref{eq:massmatrix}),
displaying the imprints of the $CBM$ mixing, whereas
%
\begin{equation}
\label{eq:corrmass}
\delta M \left(\delta,\epsilon\right) = \left(
\begin{array}{c@{\quad}c@{\quad}c}
0 & 0 & \delta
\\
0 & 0 & 0
\\
\delta& 0 & \epsilon
\end{array}
\right)
\end{equation}
modulates the deviations from the reflection symmetry. This matrix is written
in terms of two breaking parameters,
$\delta= m_{e\tau}-m_{e\mu}^{*}$ and
$\epsilon=m_{\tau\tau} - m_{\mu\mu}^{*}$, where the mass matrix entries
are now computed from
$M_{\nu}=U_{PMNS}{\mathrm{diag}}(m_{1},m_{2},m_{3}) U_{PMNS}^{
\mathrm{T}}$ by using the Eq.~(\ref{eq:newpmns}). The novelty of our approach
is then related to the fact of considering an underlaying reflection symmetry
instead of the permutation symmetry in \cite{Rivera2}, which, in addition
to our particular choice of the $PMNS$ matrix, lead to different expressions
for the breaking parameters. It is worth noting that the corrected
$CBM$ mixing matrix breaks explicitly the equalities
$m_{\tau\tau} = m_{\mu\mu}^{*}$ and $m_{e\tau}=m_{e\mu}^{*}$,
thus, the
most advantageous texture to parametrize the correction matrix, with the
minimum number of parameters, is as in Eq.~(\ref{eq:corrmass}). However,
the texture of the correction mass matrix could be written in a different
way, for instance, by locating the breaking parameters in the
$\delta M_{\mu\mu}$, $\delta M_{e \mu}$ and $\delta M_{\mu e}$ entries,
but their values would not change if the modulus of these parameters is
taken. We can also define the two dimensionless parameters in the following
way
%
\begin{eqnarray}
\label{deltaepsilon}
\hat{\delta}\equiv\frac{\delta}{m_{e\mu}^{*}} &=& \frac{-2}{1- i C}~,
\nonumber
\\
\hat{\epsilon}\equiv\frac{\epsilon}{m_{\mu\mu}^{*}} &=&
\frac{-2}{1- i \tilde C}~.
\end{eqnarray}
It is direct to show that $C$ and $\tilde C$ are functions that can be
expressed as
%
\begin{eqnarray}
\label{Cfunction}
C &=&
\frac{m_{3} c_{\beta_{2}} s_{13} - i m_{1} c_{12}
(s_{12}+ic_{12}s_{13})+i m_{2} c_{\beta_{1}}
s_{12}(c_{12}-is_{12}s_{13})}{m_{3} s_{\beta_{2}} s_{13} + i m_{2}
s_{\beta_{1}} s_{12} (c_{12}-i s_{12}s_{13})}~,
\nonumber
\\
\tilde C &=&
\frac{m_{3} c_{\beta_{2}} c_{13}^{2} + m_{1}
(s_{12}+ic_{12}s_{13})^{2} + m_{2} c_{\beta_{1}}
(c_{12}-is_{12}s_{13})^{2}}{m_{3} s_{\beta_{2}}c_{13}^{2} + m_{2}
s_{\beta_{1}} (c_{12}-is_{12}s_{13})^{2}}~,
\end{eqnarray}
where $s_{\beta_{i}} = \sin\beta_{i}$ and $c_{\beta_{i}} = \cos
\beta_{i}$, for $i=1,2$, and $m_{1,2,3}$ the
absolute values of neutrino masses. We can see from Eq.~(\ref{deltaepsilon})
that both parameters display the expected behavior as they depend on the
Majorana $CP$ phases through the functions in Eq.~(\ref{Cfunction}). It
is straightforward to show that in the limit case where
$\beta_{1}=\beta_{2}=0$ such parameters vanish, restoring the symmetric
pattern in Eq.~(\ref{eq:massmatrix}). Hence, the size of the breaking of
the reflection symmetry in the neutrino mass matrix is directly linked
to the non-vanishing values of the Majorana phases. Small deviations from
the symmetric pattern could be investigated by demanding
$|\hat\delta|,|\hat\epsilon| \lesssim1$ for the mass matrix, which
is also called a slight or soft breaking of the $\mu-\tau$ symmetry
\cite{Xing:2015fdg,Xing:2022uax}, and may help to resolve possible values
of Majorana $CP$ phases. In addition, functions $C$ and $\tilde C$ also
show a dependence on the three neutrino absolute masses. We can turn these
expressions in terms of the two squared mass differences, according to
the neutrino mass ordering, and the lightest neutrino mass, such that some
particular cases should be analyzed.

\section{Numerics}
\label{sec:results}

Before presenting a full numerical analysis, let us consider, in the following,
a semi-analytical approach and divide our discussion according to the neutrino
mass ordering. For our numerical evaluations, we will follow the results
in \cite{Esteban:2020cvm}, but the same conclusions are obtained for other
data sets.

\medskip

\noindent
\textbf{Normal Ordering}
\medskip

In the case of the normal ordering ($NO$), we can adopt the approximate
relations
$|m_{1}|\ll|m_{2}| \approx\sqrt{\Delta m^{2}_{SOL}} \ll m_{3}
\approx\sqrt{\Delta m^{2}_{ATM}}$, being $m_{1}=m_{0}$ the lightest neutrino
mass, which we leave as a free parameter. The squared mass differences
and the non-fixed angles are given at $1\sigma$ by
$\Delta m^{2}_{SOL} = (7.42^{+0.21}_{0.20})\times10^{-5}\mbox{ eV}^{2}$,
$\Delta m^{2}_{ATM} = (2.510^{+0.027}_{-0.027})\times10^{-3}\mbox{ eV}^{2}$,
$\sin^{2} \theta_{12}=(3.04^{+0.12}_{-0.12})\times10^{-1}$ and
$\sin^{2} \theta_{13} = (2.246^{+0.062}_{-0.062})\times10^{-2}$
\cite{Esteban:2020cvm}.

\begin{figure}
\includegraphics[scale=0.38]{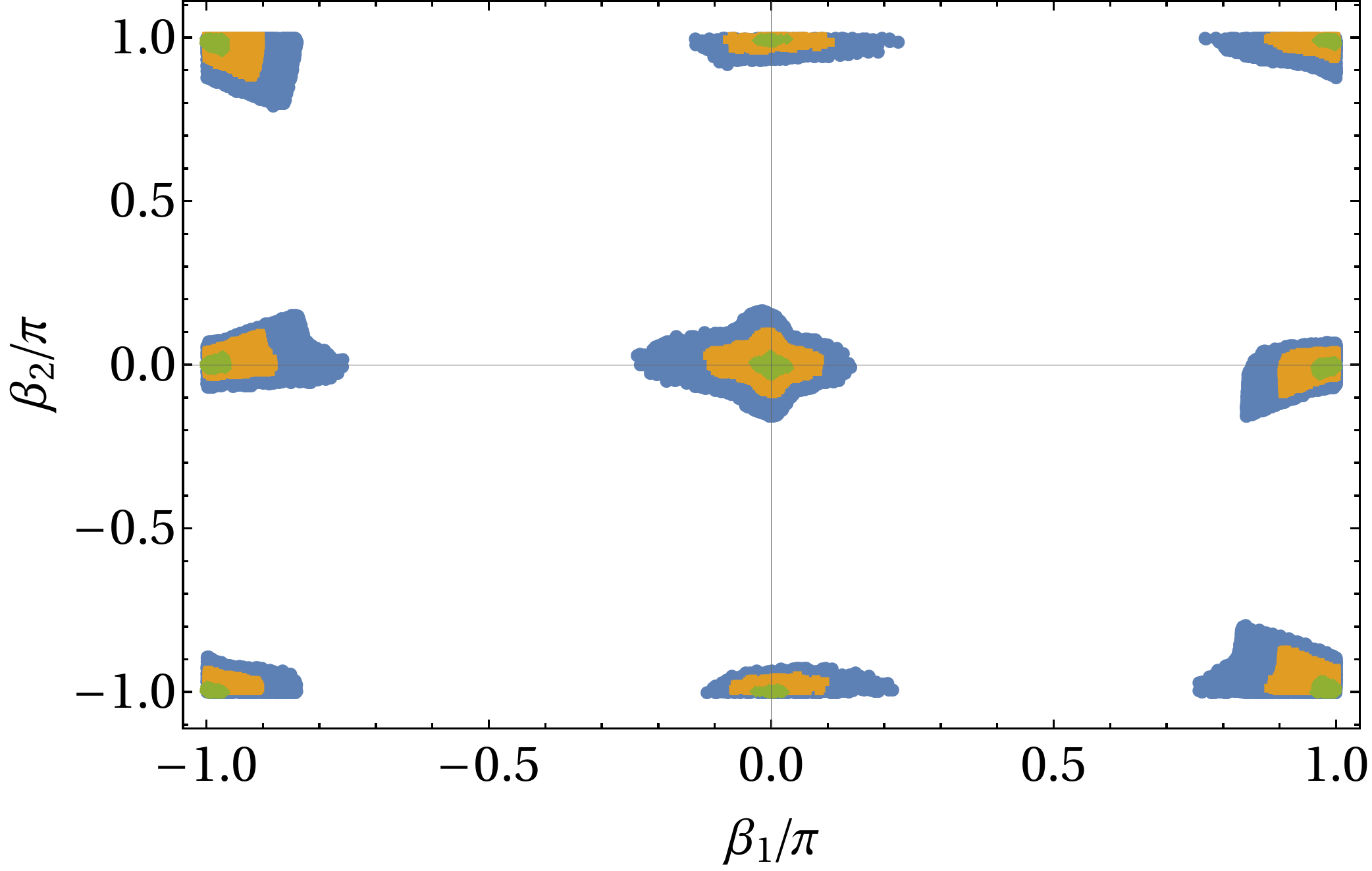} 
\includegraphics[scale=0.38]{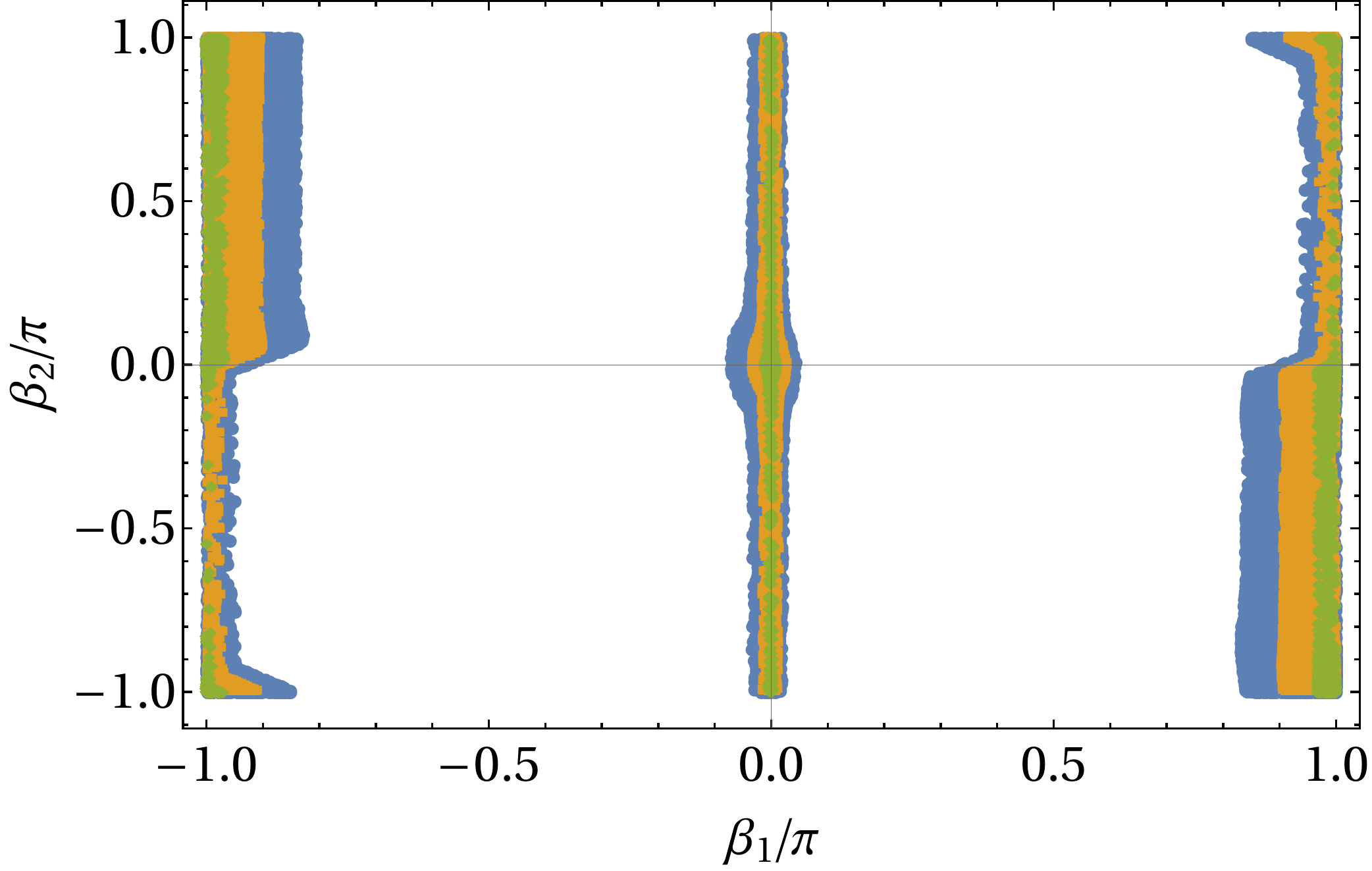}
\caption{Allowed regions of Majorana phases for $NO$ (left) and $IO$ (right).
Small, medium, and large regions (green, orange, blue) correspond to Max$[|
\hat\delta| , |\hat\epsilon|] \lesssim0.1, ~0.3, ~0.5$, respectively.}%
\label{fig:betas}
\vspace*{-4pt}
\end{figure}

A direct inspection in Eq.~(\ref{Cfunction}) shows that $m_{1}$ gives minor
contributions to $C$ and $\tilde C$ as it is suppressed in the $NO$. Hence,
it will not play a crucial role in determining $\hat\delta$ and
$\hat\epsilon$, and can be safely neglected in a first approximation.
We can also see that, for example, in the special case where
$\beta_{1} = \beta_{2} = \beta$, we obtain
$\hat{\delta}\approx\hat{\epsilon} \approx-2(1- i \cot\beta)^{-1}$.
It is also easy to show that the combination where one of the Majorana
phases is fixed to zero leaves a similar expression for the non-zero phase.
Hence, we can observe that values of $\beta$ in the $CP$ conserving limit
($0, \pm\pi$) recover the symmetry in the neutrino mass matrix of Eq.~(\ref{eq:corrected}). On the other hand, the maximal $CP$ violation case
($\pm\pi/2$) produces large deviations from the symmetric scenario since
$|\hat\delta|\approx|\hat{\epsilon}| \approx2$. We should then expect
that, in the $NO$, if small deviations from the symmetric mass matrix are
demanded, the Majorana phases should remain near the $CP$ conserving values.
This can be observed in the left plot in Fig.~\ref{fig:betas}, which are
obtained for the full expressions.

\medskip

\noindent
\textbf{Inverted Ordering}

\medskip

For the inverted ordering ($IO$), we have the following relations:
$|m_{1}| \approx\sqrt{\Delta m_{ATM}^{2}}$,
$|m_{2}| \approx\sqrt{\Delta m _{SOL}^{2} + \Delta m_{ATM}^{2}} \gg m_{3}$.
In this case, $m_{3} = m_{0}$ is the lightest neutrino mass, and
$\Delta m^{2}_{SOL} = (7.42^{+0.21}_{0.20})\times10^{-5}\mbox{ eV}^{2}$,
$\Delta m^{2}_{ATM} = -(2.490^{+0.026}_{-0.028}) \times10^{-3}\mbox{ eV}^{2}$,
$\sin^{2} \theta_{12}=(3.04^{+0.13}_{-0.12})\times10^{-1}$ and
$\sin^{2} \theta_{13} = (2.241^{+0.074}_{-0.062})\times10^{-2}$
\cite{Esteban:2020cvm}.

We observe in this case that, neglecting the lightest neutrino mass, the
breaking parameters take the approximated form
%
\begin{eqnarray}
\label{eq:deltaepsilonIO}
\hat{\delta} &\approx&
\frac{-2 i s_{\beta_{1}} e^{-i\beta_{1}}}{1 - e^{-i\beta_{1}} \left
( \frac
{c_{12}s_{12}+ic_{12}^{2}s_{13}}{c_{12}s_{12}-is_{12}^{2}s_{13}} \right)}
+ {\mathcal{O}} \left(
\sqrt{\frac{\Delta m_{SOL}^{2}}{\Delta m_{ATM}^{2}} }~ \right)~,
\nonumber
\\
\hat{\epsilon} &\approx&
\frac{-2 i s_{\beta_{1}} e^{-i\beta_{1}}}{1 + e^{-i\beta_{1}} \left
( \frac{s_{12}+ic_{12}s_{13}}{c_{12}-is_{12}s_{13}}\right)^{2} }
+ {\mathcal{O}} \left(
\sqrt{ \frac{\Delta m_{SOL}^{2}}{\Delta m_{ATM}^{2}} }~\right) ~.
\end{eqnarray}
Within this approximation, we observe that the breaking parameters do not
have a strong dependence on $\beta_{2}$, such that we should expect that
this phase will not be restricted in our analysis of the values of
$\hat\delta$ and $\hat\epsilon$. It is also important to note that
the squared mass differences, and the mixing angles, will play a sub-leading
role in the determination of these parameters, with the main contribution
coming from $\beta_{1}$. Hence, we can verify that the exact symmetry
limit is related to $CP$ conserving values ($0,\pm\pi$) of
$\beta_{1}$, while deviations from the symmetric pattern are then governed
in this limit by specific values of $\beta_{1}$. For instance, for
$\beta_{1} \approx0.1 ~(\sim6^{\circ})$ we obtain
$|\hat\delta| \approx|\hat\epsilon| \approx0.2$. On the other hand,
a maximal value of this phase $(\beta_{1} = \pm\pi/2)$ leads to a large
deviation from the symmetric pattern, i.e.,
$|\hat\delta| \approx|\hat\epsilon| \approx2$. The full numerical
analysis is resumed in the right plot of Fig.~\ref{fig:betas}.

\medskip

\noindent
\textbf{Degenerate Ordering}

\medskip

Finally, we can also analyze the degenerate ordering, where absolute masses
are the same order $|m_{1}|\approx|m_{2}| \approx|m_{3}|$. In this case,
the breaking parameters show no strong dependence on the mass, but the
limit is restricted to large values of the lightest neutrino mass, which
is almost excluded from cosmological observations \cite{Ade:2015xua}. In
this limit, the dependence of the breaking parameters on the mixing angles
and $CP$ phases can be directly obtained from Eq.~(\ref{deltaepsilon}).
As in the previous cases, the $CP$ conserving limit is related to the null
values of the breaking parameters. A numerical inspection shows that, when
we take $\beta_{1}=\beta_{2}=\beta$, these parameters will have a similar
form to the first terms in Eq.~(\ref{eq:deltaepsilonIO}), leading to
$|\hat\delta| , |\hat\epsilon| \gtrsim1$ for
$(\beta= \pm\pi/2)$, which should be excluded in the search of small
deviations. This pattern is also present in both $NO$ and $IO$, but leaves
open questions about other different combinations in the values of
$CP$ phases consistent with
$|\hat\delta| , |\hat\epsilon| \lesssim1$.

Our previous analysis may serve to confirm the expectations that the
$CP$ conserving values of Majorana phases are linked to the symmetric structure
of neutrino mass matrix, but also shows that we should investigate other
possible combinations of $CP$ phases which may lead to small deviations
from this symmetric pattern. To this aim, a full numerical analysis is
also mandatory, considering different values of $m_{0}$, the mass orderings,
and the deviations in the mixing angles from their central values. We show
in Fig.~\ref{fig:betas} the different allowed regions of Majorana phases
for different values of breaking parameters. In these plots, we run the
lightest mass from zero to $0.4\mbox{ eV}$, in both mass orderings, and allowed
the mixing angles to vary within the $3\sigma$ interval. We show, for
comparison, three different cases:
$|\hat\delta| , |\hat\epsilon| \lesssim0.1$,
$|\hat\delta| , |\hat\epsilon| \lesssim0.3$, and
$|\hat\delta| , |\hat\epsilon| \lesssim0.5$. We can observe that other
different combinations of Majorana phases are obtained, out of the
$CP$ conserving limit, which leads to small (or slight) deviations from
the symmetric mass matrix. As we could anticipate, the size of the allowed
region depends on the selected values of the breaking parameters. It is
worth noting that, while for the $NO$ both of the $CP$ phases are restricted
to lie near the $CP$ conserving values, in the $IO$, the phase
$\beta_{2}$ can reach a maximal value even for very small deviations.
Such regions, and also the differences between different mass orderings,
could be confronted with some physical observables as in the case of
neutrinoless
double beta decay.

\begin{figure}
\includegraphics[scale=0.6]{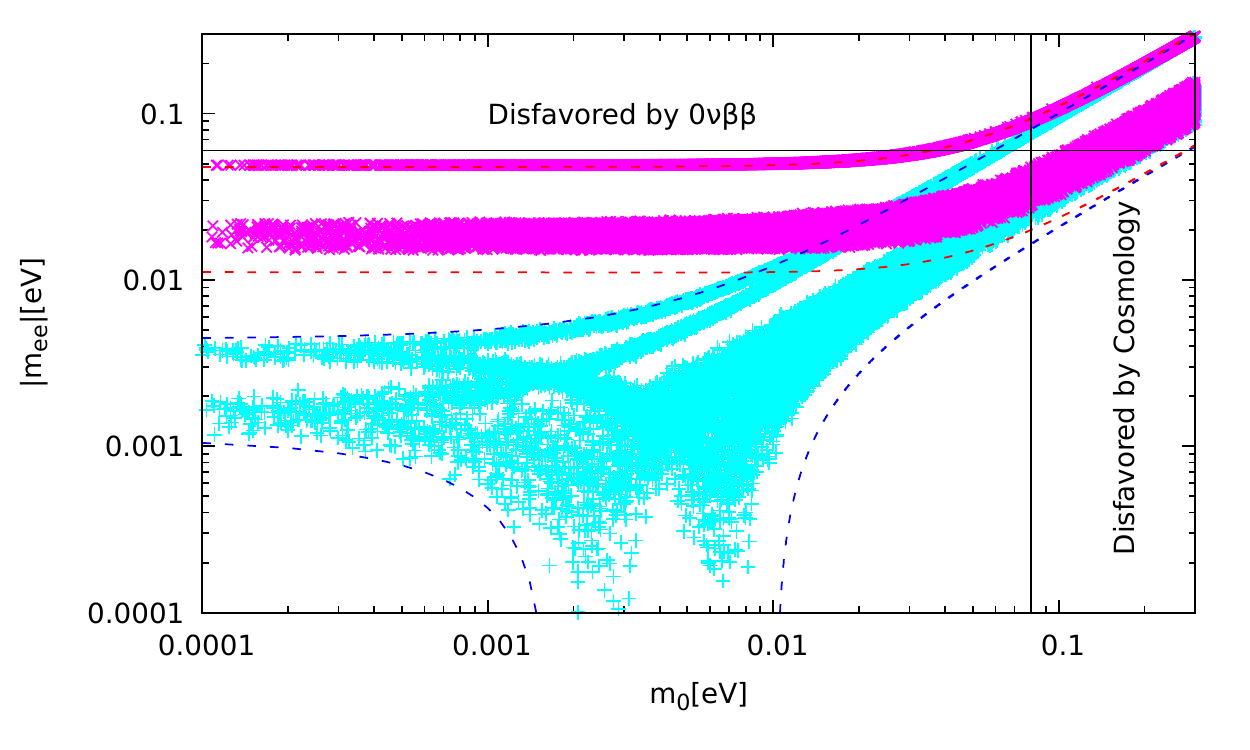} 
\includegraphics[scale=0.6]{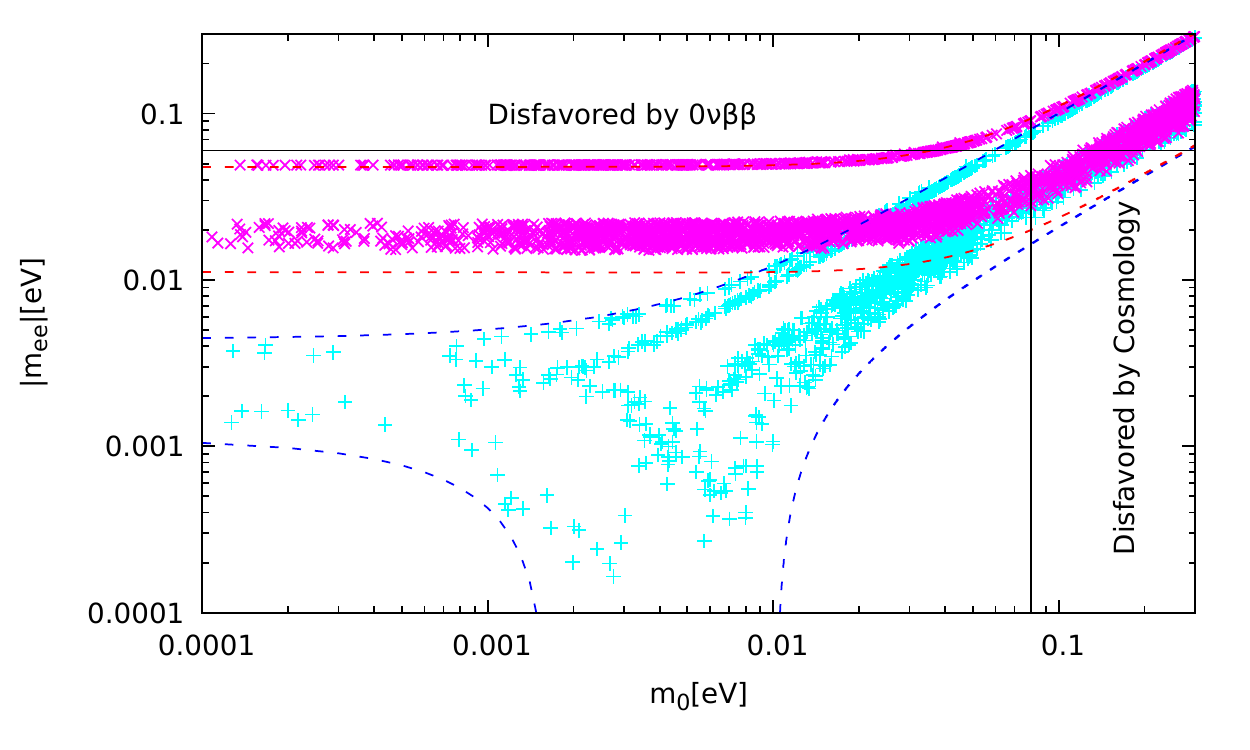}
\caption{Allowed regions of $|m_{ee}|$ for different lightest neutrino mass in
the $NO$ (cyan) and $IO$ (magenta). Regions delimited by dotted lines represent
the full region of $NO$ and $IO$ for nonrestricted $CP$ phases. Regions excluded
by cosmology and $0\nu\beta\beta$ experiments are delimited by solid lines
\cite{Ade:2015xua,Auger:2012ar,Albert:2014awa,Gando:2012zm}. Left plot is for
Max$[|\hat\delta| , |\hat\epsilon|] \lesssim0.5$, and right plot for
Max$[|\hat\delta| , |\hat\epsilon|] \lesssim0.1$.}%
\label{fig:doublebeta}
\vspace*{-4pt}
\end{figure}

In Fig.~\ref{fig:doublebeta}, we plot the matrix element of neutrinoless
double beta decay ($|m_{ee}|$) depending on the lightest neutrino mass.
We can observe that the combinations of Majorana phases obtained from the
restrictions in $|\hat\delta|$ and $|\hat\epsilon|$ give specific regions
in $|m_{ee}|$, which may be compared with forthcoming experimental observations.
For comparison, we show in the left plot the allowed regions for the condition
$|\hat\delta| , |\hat\epsilon| \lesssim0.5$, which present a slight
reduction compared to the case of totally free Majorana phases. In the
right plot of Fig.~\ref{fig:doublebeta}, we present the region corresponding
to the restriction $|\hat\delta| , |\hat\epsilon| \lesssim0.1$. We
observe, in this case, that some specific values of $|m_{ee}|$ would not
be compatible with a slightly broken symmetry in the mass matrix.

We have centered our analysis on an indirect determination of the Majorana
phases, which seem to be very restricted when a $\mu-\tau$ reflection
symmetry is demanded in the mass matrix. These regions may be extended
when this last requirement is relaxed, giving place to the possibility
of having different combinations of values consistent with small deviations.
In the $IO$, it is also possible to have one of the Majorana phases with
a maximal value even for an exact symmetry in the mass matrix. Some of
our results could be of theoretical interest in the search for models of
neutrino masses and mixings.

\section{Summary}
\label{sec:conclusions}
\vspace*{-3pt}
A $CBM$ mixing matrix is of great theoretical interest as it is related
to some discrete symmetries, which could help to understand the pattern
of masses and mixings. We have shown, from a phenomenological approach,
that $CP$ nonconserving values of the Majorana phases may break the symmetric
structure of the neutrino mass matrix, regardless of the symmetry in the
mixing matrix. By modulating the deviations with two breaking parameters,
we found different combinations of Majorana phases consistent with small
departures from the symmetric scenario, which could be of interest in the
search for a perturbative treatment of the neutrino mass matrix, preserving
the $CBM$ symmetry in the $PMNS$ matrix. In addition, we have also shown
that simultaneous maximal values of Majorana phases are associated with
large deviations from the symmetric limit, but, in the $IO$, it is possible
to have one maximal Majorana phase consistent with the $\mu-\tau$ symmetry
of the mass matrix. Restricted regions were obtained for the neutrinoless
double beta decay amplitude, depending on the size of the deviations, which
could be confronted with forthcoming results.

   
\section*{Acknowledgements}

 The authors acknowledge funding from  \textit{Division General de Investigaciones} (DGI) of the Santiago de Cali University under grant 935-621121-3068.


\bibliographystyle{utphys}
\vspace*{-10pt}

\providecommand{\href}[2]{#2}\begingroup\raggedright\endgroup


\end{document}